\documentclass{revtex4}
\usepackage{graphicx}
\usepackage{dcolumn}
\usepackage{amsmath}
\newcommand{\md}[0]{\mbox{d}}
\begin{document}

\title{Coefficient of restitution for viscoelastic disks}
\author{Thomas Schwager}
\homepage{http://bioinf.charite.de/biophys/people/schwager/}
\email{thomas.schwager@charite.de}
\affiliation{Charit{\'e} Berlin -- Centrum f{\"u}r muskuloskeletale Chirurgie, Augustenburgerplatz 1, D-13353 Berlin, Germany}
\date{\today}

\begin{abstract}
  The dissipative collision of two identical viscoelastic disks is studied. By using a known law for the elastic part of the interaction force and the viscoelastic damping model an analytical solution for the coefficient of restitution shall be given. The coefficient of restitution depends significantly on the impact velocity. It approaches one for small velocities and decreases for increasing velocities.
\end{abstract}
\maketitle

\section{Introduction}

The coefficient of restitution is an important means to characterize the damping properties of granular particles. It is defined as
\begin{equation}
  \epsilon=\frac{g^{\prime}}{g}
\end{equation}
with $g$ being the absolute value of the normal component of the relative velocity before and $g^{\prime}$ the value after the collision. 
In most analytical and numerical studies of the behaviour of dilute or fluidized granular systems a constant coefficient of restitution was used (see, e.g., \cite{Taguchi:1992PRL,GoldhirschZanetti:1993,DuLiKadanoff:1994,GoldshteinShapiro:1995,NoijeErnst:1998,BritoErnst:1998} and many more). In experiments on (3d) spheres, however, it has been found that this coefficient is not a constant but depends significantly on impact velocity \cite{HatzesBridgesLin:1988}. For spheres there is a theory based on first principles which predicts this velocity-dependence \cite{SchwagerPoeschel:1998,RamirezPoeschelBrilliantovSchwager:1999}. The aim of the present paper is to use the methods successfully applied to spheres for the description of the collisional properties of disks.

If we assume particles whose conservative interaction part is linear with respect to the mutual compression and whose dissipative interaction part is linear with respect to the relative velocity the coefficient of restitution would indeed be constant \cite{SchaeferDippelWolf:1995}. This can be very useful if one wishes to compare results from Molecular Dynamics (or Discrete Elements) simulations with the predictions from theory based on $\epsilon=$const. Apart from these practical considerations there are authors who use disks as a real world examples of particles interacting via a linear force law (see, e.g., \cite{Laetzeletal:2000,GoldenbergGoldhirsch:2004}). However, it is known for several decades that the contact force law for colliding disks differs significantly from a linear law \cite{Engel}:

\begin{equation}
  \label{eq:implicitelaw}
  \xi = \frac{F_{\rm el}}{\pi Y}\left[\ln\frac{4\pi RY}{F_{\rm el}}-1-\nu\right]
\end{equation}
Here $\xi=2R-\left|\vec{r}_1-\vec{r}_2\right|$ is the compression of the particles of radius $R$ at positions $\vec{r}_{1/2}$. The material properties are characterized by the Young modulus $Y$ and the Poisson ratio $\nu$. One should note that in the present context ''disk'' means infinetely long cylinder. Above expression is, thus, only approximately valid for disks of finite thickness (or cylinders of finite length). For too thin disks the (implicit) force law \eqref{eq:implicitelaw} may become wrong. Eq. \eqref{eq:implicitelaw}, can be solved for the contact force $F_{\rm el}$ yielding
\begin{equation}
  \label{eq:explicitlaw}
  F_{\rm el}=-\frac{\pi Y\xi}{\displaystyle W_0\left(-\frac{e^{1+\nu}}{4R}\xi\right)}\approx-\frac{\pi Y\xi}{\displaystyle\ln\frac{e^{1+\nu}\xi}{4R}}~.
\end{equation}
The function $W_0$ is the zeroth Lambert-$W$-function. It has two real and infinitely many complex branches. One of them vanishes linearly as the argument vanishes, the others diverge as the logarithm of the modulus of the argument. The first (vanishing) branch is irrelevant since it would yield a finite force for infinitesimal deformation. The relevant details of $W_0$ are summarized in appendix \ref{appendix:lambert}. The force law \eqref{eq:explicitlaw} was used to study the coefficient of restitution of colliding disks where energy was dissipated via excitation of internal degrees of freedom of the disks, i.e., by excitation of vibrational modes \cite{GerlZippelius:1999}. This dissipation mechanism will be neglected here. 

In this article we assume viscoelastic damping of the deformed particle material. For this damping mechanism the material of contacting particles is assumed to be locally linear, i.e. the stress tensor depends linearly on both the deformation tensor and the deformation rate tensor. If the impact velocity is small enough this assumption is reasonable. For higher velocities other dissipation mechanisms, like plastic deformation or the already mentioned excitation of vibrational modes have to be taken into account. The authors of \cite{BrilliantovSpahnHertzschPoeschel:1994} used the viscoelastic model to study the contact of spheres and found that the dissipative force component $F_{\rm diss}$ is related to the conservative force $F_{\rm el}$ via
\begin{equation}
  \label{eq:dissforceprincipalstructure}
  F_{\rm diss}=A\,\dot{\xi}\,\frac{\partial F_{\rm el}}{\partial\xi}~,
\end{equation}
where (in 3d) $F_{\rm el}$ is the Hertz contact force. Subsequently the above expression Eq. \eqref{eq:dissforceprincipalstructure} was derived using a crystal mechanical approach \cite{MorgadoOppenheim:1997}. 
The parameter $A$ is proportional to the viscous relaxation time of the particles and depends only on the material constants and the geometry, such as the radius of the particles. Due to the general nature of the argumentation from \cite{BrilliantovSpahnHertzschPoeschel:1994} it can be applied to the colliding disks problem too. Inserting the force law, Eq. \eqref{eq:explicitlaw}, into Eq. \eqref{eq:dissforceprincipalstructure} yields
\begin{equation}
  \label{eq:dampinglaw}
  F_{\rm diss}= -\frac{\pi AY\dot{\xi}}{1+W_0\left(\displaystyle-\frac{e^{\nu+1}}{4R}\xi\right)}~.
\end{equation}
With Eq. (\ref{eq:dissforceprincipalstructure}) the equation of motion reads
\begin{eqnarray}
  \label{eq:eqnofmotion1}
  m_{\rm eff}\ddot{\xi}+F_{\rm el}+A\,\dot{\xi}\,\frac{\partial F_{\rm el}}{\partial \xi}&=&0\\
  \xi(0)&=&0\\
  \dot{\xi}(0)&=&g
  \label{eq:eqnofmotion3}
\end{eqnarray}
where repulsive forces are positive. We introduce rescaled variables for length, force and time
\begin{eqnarray}
  \label{eq:scalex}
  x&=&\frac{e^{1+\nu}}{4R}\xi\\
  \tilde{F}_{\rm el}&=&\frac{\displaystyle e^{1+\nu}}{4\pi RY}F_{\rm el}\\
  \tau&=&\sqrt{\frac{\pi Y}{m}}t
\end{eqnarray}
and define the scaled velocity $v$ and the scaled damping parameter $\alpha$ by 
\begin{eqnarray}
  v = \frac{\md x}{\md\tau} &=&g\frac{\displaystyle e^{1+\nu}}{4R}\sqrt{\frac{m}{\pi Y}}\\
  \alpha &=& A\sqrt{\frac{\pi Y}{m}}\label{eq:scalealpha}~.
\end{eqnarray}
Introducing Eqs. \eqref{eq:scalex}-\eqref{eq:scalealpha} into \eqref{eq:eqnofmotion1}-\eqref{eq:eqnofmotion3} we obtain
\begin{eqnarray}
  \label{eq:eqnofmotion}
  \ddot{x}+\tilde{F}_{\rm el}+\alpha\dot{x}\frac{d\tilde{F}_{\rm el}}{dx}&=&0\\
  x(0)&=&0\\
  \dot{x}(0)&=&v~,
\end{eqnarray}
There is a simple interpretation of $v$ and $\alpha$. We can rewrite $m=\rho~\pi R^2$ (with $\rho$ being the material density) and find $v\sim g\sqrt{\rho/Y}$. In a rough approximation one can estimate the speed of sound by $c\sim\sqrt{Y/\rho}$ thus 
\begin{equation}
  v\sim \frac{g}{c}~.
\end{equation}
By a similar argumentation we find that $\alpha\sim Ac/R$, i.e., $\alpha$ can be estimated by the viscous relaxation time (which is the meaning of the damping parameter $A$) divided by the time a sound wave needs to travel through the cross-section of the particle. Obviously both $v$ and $\alpha$ must be small compared to unity. 

\begin{figure}[htbp]
  \centerline{\includegraphics[width=8cm,clip]{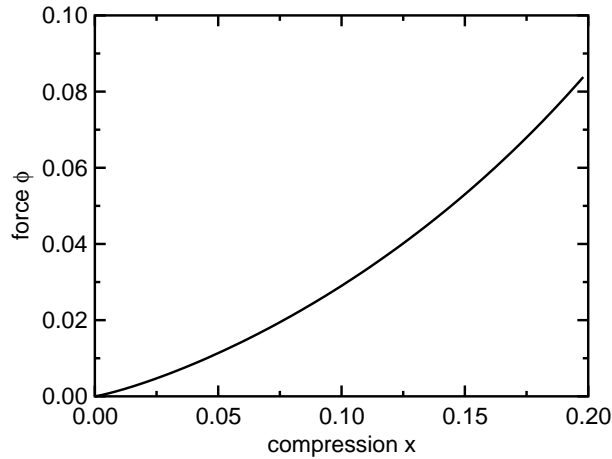}}
  \caption{The rescaled elastic force $\tilde{F}_{\rm el}(x)$ as a function of the rescaled compression $x$.}
  \label{fig:cylforce}
\end{figure}

The implicite elastic force law in the rescaled variables does not contain any dependence on material parameters 
\begin{equation}
  \label{eq:scaledimplicitelaw}
  x=-\tilde{F}_{\rm el}\ln\tilde{F}_{\rm el}~.
\end{equation}
The explicit solution of Eq. (\ref{eq:scaledimplicitelaw}) is
\begin{equation}
  \label{eq:scaledexplicitlaw}
  \tilde{F}_{\rm el}=-\frac{x}{W_0(-x)}
\end{equation}
and can alternatively be obtained substituting the normal variables by the rescaled ones in Eq. \eqref{eq:explicitlaw}. The rescaled force as a function of the rescaled compression is given in Fig. \ref{fig:cylforce}.

The dissipative part of the interaction force (see Eq. \eqref{eq:dissforceprincipalstructure}) is
\begin{equation}
  \tilde{F}_{\rm diss}= \alpha\dot{x}\frac{d\tilde{F}_{\rm el}}{dx}=-\alpha\frac{\dot{x}}{1+\ln\tilde{F}_{\rm el}}=-\alpha\frac{\dot{x}}{1+W_0(-x)}~.
\end{equation}

In a straightforward approach the coefficient of restitution can be found by integrating Newton's equation of motion and determining the trajectory of the particles $x(\tau)$, the duration of the collision $\tau_c$ and, eventually, $\epsilon=-\dot{x}(\tau_c)/v$. Here, however, we use a different approach due to mathematical difficulties resulting from the complicated form of the force law. 

In the next section \ref{sec:balance} an approach which doesn't rely on the explicit knowledge of the trajectory is presented. The method allows to compute the dominant part of the energy loss during the collision. As a further advantage we do not need to know explicit expressions for the damping force, we only need to know its principal structure (\ref{eq:dissforceprincipalstructure}), which is especially useful for more complicated force laws. In section \ref{sec:consistency} a method to further improve the obtained result will be presented.

\section{Energy balance}\label{sec:balance}
\noindent
The coefficient of restitution and the total energy loss during a collision are related by 
\begin{equation}
  1-\epsilon^2=2\frac{\Delta E}{v^2}~.
\end{equation}
The value of $\Delta E$ up to linear order in the damping parameter $\alpha$ can be found {\em without} explicitly knowing the trajectory of the particle. This procedure was successfully employed before to compute the coefficient of restitution for colliding spheres \cite{KuwabaraKono:1987,RamirezPoeschelBrilliantovSchwager:1999,SchwagerPoeschel:1998,HayakawaKuninaka:2004}. We introduce the interaction potential $\Phi$ and rewrite the equation of motion (\ref{eq:eqnofmotion})
\begin{equation}
  \label{eq:ebalancelong}
  \frac{\md}{\md\tau}\left(\frac{\dot{x}^2}{2}+\Phi\right) = -\alpha\dot{x}^2\frac{\md \tilde{F}_{\rm el}}{\md x}~,
\end{equation}
introducing the potential $\Phi$ via 
\begin{equation}
  F_{\rm el}=\frac{\md\Phi}{\md x}\equiv\Phi^{\prime}~.
\end{equation}
Note that the custumary negative sign is missing in the above definition as we count repulsive forces positive. The potential energy shall vanish for vanishing deformation. The term in brackets on the left hand side of Eq. (\ref{eq:ebalancelong}) is the total energy,
\begin{equation}
  \label{eq;totalenergy}
  E=\frac{\dot{x}^2}{2}+\Phi
\end{equation}
(with initial value $E_0=v^2/2$). Thus we can rewrite Eq. (\ref{eq:ebalancelong}) and obtain the energy balance equation 
\begin{equation}
  \label{eq:ebalanceshort}
  \frac{\md E}{\md\tau}=-\alpha\dot{x}^2\frac{\md\tilde{F}_{\rm el}}{\md x}=-\tilde{F}_{\rm diss}\dot{x}~.
\end{equation}
For the total energy loss during the collision we thus find
\begin{equation}
  \label{eq:totalenergyloss}
  \Delta E=-\alpha\int\limits_0^{\tau_c}\md\tau~ \dot{x}^2\frac{\md\tilde{F}_{\rm el}}{\md x}~,
\end{equation}
where $t_c$ is the duration of the collision. To actually evaluate this integral we would now need expressions for the trajectory of the particles. However, for small damping the trajectory will be close to the undamped trajectory $x_0$, i.e. $x(\tau)=x_0(\tau)+\alpha x_1(\tau)+{\cal O}(\alpha^2)$, where the first order correction term $x_1$ is well behaved for all practical collision problems. Inserting this ansatz into Eq. (\ref{eq:totalenergyloss}) we find that we can estimate the energy loss in linear order of $\alpha$ by evaluating the integral along the {\em undamped} trajectory instead of the correct damped one. We thus find 
\begin{equation}
  \label{eq:totalenergylosssimplified}
  \Delta E=-\alpha\int\limits_0^{\tau_c^0}\md\tau ~\dot{x}_0^2\frac{\md\tilde{F}_{\rm el}}{\md x} + {\cal O}\left(\alpha^2\right)~,
\end{equation}
where $\tau_c^0$ is the duration of the undamped collision. The force $\tilde{F}_{\rm el}$ has to be evaluated at the positions of the undamped trajectory. This approximation allows us to determine $\Delta E$ without explicitely knowing the trajectory. We split the integral into two parts
\begin{equation}
  \Delta E = -\alpha\int\limits_0^{\tau_c^0/2}\md\tau \dot{x}_0^2\frac{\md\tilde{F}_{\rm el}}{\md x} - \alpha\int\limits_{\tau_c^0/2}^{\tau_c^0}\md\tau \dot{x}_0^2\frac{\md\tilde{F}_{\rm el}}{\md x} + {\cal O}\left(\alpha^2\right)~,
\end{equation}
where $\tau_c^0/2$ is the time of maximum compression, where the particles are momentarily at rest and start to seperate again. Since the trajectory is symmetrical with respect to this time the second part of the integral is exactly equal to the first. 
\begin{equation}
  \Delta E = -2\alpha\int\limits_0^{\tau_c^0/2}\md\tau \dot{x}_0^2\frac{\md\tilde{F}_{\rm el}}{\md x} + {\cal O}\left(\alpha^2\right)
\end{equation}
Since in the interval $(0,\tau_c^0/2)$ the force is a monotonous function of time we can replace the time integral by an integral over the (elastic) contact force, i.e.,
\begin{equation}
  \md\tau\dot{x}\frac{\md\tilde{F}_{\rm el}}{\md x} = \md\tilde{F}_{\rm el}~.
\end{equation}
Furthermore we express the velocity of the undamped problem in terms of the potential energy, which gives
\begin{eqnarray}
  \dot{x}_0&=&\sqrt{2(E-\Phi)}=\sqrt{v^2-2\Phi}~,
\end{eqnarray}
with $v$ being the (scaled) impact velocity. Therefore, 
\begin{equation}
  \Delta E = -2\alpha\int\limits_0^{\tilde{F}_{\rm max}}\md\tilde{F}_{\rm el}\sqrt{v^2-2\Phi} + {\cal O}(\alpha^2)~.
\end{equation}
We express the potential energy in terms of the force, i.e., $\Phi=\Phi(\phi)$. 
\begin{eqnarray}
  \tilde{F}_{\rm el}&=&\frac{\md\Phi}{\md x}=\frac{\md\Phi}{\md\tilde{F}_{\rm el}}\frac{\md\tilde{F}_{\rm el}}{\md x}\\
  \frac{\md\Phi}{\md\tilde{F}_{\rm el}}&=&\tilde{F}_{\rm el}\left(\frac{\md \tilde{F}_{\rm el}}{\md x}\right)^{-1}
\end{eqnarray}
From Eq. (\ref{eq:scaledimplicitelaw}) we find $\md \tilde{F}_{\rm el}/\md x=-1/\left(1+\ln\tilde{F}_{\rm el}\right)$ and thus
\begin{equation}
  \frac{\md\Phi}{\md\tilde{F}_{\rm el}}=-\tilde{F}_{\rm el}\left(1+\ln\tilde{F}_{\rm el}\right)~.
\end{equation}
With $\Phi(x=0)=0$ we find
\begin{equation}
  \label{eq:Phiofphi}
  \Phi=-\frac{\tilde{F}_{\rm el}^2}{4}\left(1+2\ln\tilde{F}_{\rm el}\right)~.
\end{equation}
Therefore
\begin{equation}
  \Delta E = -2\alpha v\int\limits_0^{\tilde{F}_{\rm max}}d\phi\sqrt{1 + \frac{\tilde{F}_{\rm el}^2\left(1+2\ln\tilde{F}_{\rm el}\right)}{2v^2}} + {\cal O}(\alpha^2)~.
\end{equation}
The maximum elastic force $\tilde{F}_{\rm max}$ is related to the impact velocity by inserting the energy conservation $v^2=2\Phi$ into Eq. (\ref{eq:Phiofphi})
\begin{equation}
  \label{eq:phimaximplicit}
  2v^2=-\tilde{F}_{\rm max}^2\left(1+2\ln\tilde{F}_{\rm max}\right)~.
\end{equation}
We will now use this expression to replace $v$ in the integral. The actual solution $\tilde{F}_{\rm max}(v)$ will be given at a later stage of the computation.
\begin{equation}
  \Delta E = -2\alpha v\int\limits_0^{\tilde{F}_{\rm max}}\md\phi\sqrt{1 - \frac{\tilde{F}_{\rm el}^2\left(1+2\ln\tilde{F}_{\rm el}\right)}{\tilde{F}_{\rm max}^2\left(1+2\ln\tilde{F}_{\rm max}\right)}} + {\cal O}(\alpha^2)
\end{equation}
Substituting $\tilde{F}_{\rm el}=\tilde{F}_{\rm max}\,y$ with $y$ from the interval $(0,1)$ we find
\begin{eqnarray}
  \Delta E &=& -2\alpha v\phi_{\rm max} \int\limits_0^1dy\sqrt{1 - \frac{y^2\left(1+2\ln\tilde{F}_{\rm max}+2\ln y\right)}{\left(1+2\ln\tilde{F}_{\rm max}\right)}} + {\cal O}(\alpha^2)\\
  &=& -2\alpha v\phi_{\rm max} \int\limits_0^1dy\sqrt{1 - y^2 - \frac{2y^2\ln y}{\left(1+2\ln\tilde{F}_{\rm max}\right)}} + {\cal O}(\alpha^2)\\
  &=& -2\alpha v\phi_{\rm max} \int\limits_0^1dy \sqrt{1-y^2}\sqrt{1 - \frac{2y^2\ln y}{\left(1-y^2\right)\left(1+2\ln\tilde{F}_{\rm max}\right)}} + {\cal O}(\alpha^2)\\
\end{eqnarray}
In the interval $y\in (0,1)$ the relation 
\begin{equation}
  -\frac{2y^2\ln y}{1-y^2}\le 1~.
\end{equation}
holds. For sufficiently small impact velocity $v$
we can, therefore, expand the second term in the integrand into a power series with respect to the small term $2y^2\ln y/\left[\left(1-y^2\right)\left(1+2\ln\phi_{\rm max}\right)\right]$. With $a_k$ being the Taylor coefficients of the expansion of $\sqrt{1+x}$ around $x=0$ we find
\begin{eqnarray}
  \Delta E &=& -2\alpha v\tilde{F}_{\rm max} \sum\limits_{k=0}^{\infty}(-1)^ka_k\int\limits_0^1dy \sqrt{1-y^2}\left[\frac{2y^2\ln y}{\left(1-y^2\right)\left(1+2\ln\tilde{F}_{\rm max}\right)}\right]^k + {\cal O}(\alpha^2)\\
  &=&-2\alpha v\tilde{F}_{\rm max} \sum\limits_{k=0}^{\infty}\frac{(-1)^ka_k}{\left(1+2\ln\tilde{F}_{\rm max}\right)^k}\int\limits_0^1dy \sqrt{1-y^2}\left[\frac{2y^2\ln y}{1-y^2}\right]^k + {\cal O}(\alpha^2)\\
  &=& -2\alpha v\tilde{F}_{\rm max} \sum\limits_{k=0}^{\infty}\frac{(-1)^kc_k}{\left(1+2\ln\tilde{F}_{\rm max}\right)^k} + {\cal O}(\alpha^2)~,
\end{eqnarray}
with 
\begin{equation}
  c_k=a_k\int\limits_0^1dy \sqrt{1-y^2}\left[\frac{2y^2\ln y}{1-y^2}\right]^k
  \label{eq:definitionck}
\end{equation}
For $c_1$ and $c_2$ there exist analytical expressions, the higher coefficients have to be computed numerically. The first values of $c_k$ are given in Table \ref{tab:firstvaluesckdk} at the end of next section. We now have to find an expression for $\tilde{F}_{\rm max}$ in terms of $v$. The defining equation (\ref{eq:phimaximplicit}) can be solved in closed form using again Lambert's $W$-function:
\begin{eqnarray}
  \tilde{F}_{\rm max}&=&\sqrt{\frac{2v^2}{-W_0(-2ev^2)}}\\
  1+2\ln\tilde{F}_{\rm max}&=&W_0(-2ev^2)
\end{eqnarray}
with the Euler constant $e$. Note that the function $W_0$ is negative for negative arguments. The final expression for the energy loss reads
\begin{eqnarray}
  \Delta E &=& -2\alpha v^2 \sqrt{\frac{2}{-W_0(-2ev^2)}} \sum\limits_{k=0}^{\infty}\frac{(-1)^kc_k}{\left[W_0(-2ev^2)\right]^k} + {\cal O}(\alpha^2)\\
  &=& -\sqrt{8}\,\alpha v^2 \sum\limits_{k=0}^{\infty}\frac{c_k}{\left[-W_0(-2ev^2)\right]^{k+1/2}} + {\cal O}(\alpha^2)~.\label{eq:firstorderenergyloss}
\end{eqnarray}
The coefficient of restitution can now be determined with the same accuracy, i.e., linearly in $\alpha$. We have 
\begin{eqnarray}
  \epsilon &=& \sqrt{1+\frac{\Delta E}{E}}\\
  &=& \sqrt{1+\frac{2\Delta E}{v^2}}\\
  &\approx& 1 + \frac{\Delta E}{v^2}
\end{eqnarray}
Inserting Eq. (\ref{eq:firstorderenergyloss}) we arrive at the final expression for the restitution coefficient up to linear order in the damping parameter $\alpha$. 
\begin{equation}
  \epsilon = 1 - \sqrt{8}\,\alpha\sum\limits_{k=0}^{\infty}\frac{c_k}{\left[-W_0(-2ev^2)\right]^{k+1/2}} + {\cal O}(\alpha^2)
  \label{eq:firstorderepsilon}
\end{equation}
For {\em very} small velocities this expression can be approximated as 
\begin{eqnarray}
  1-\epsilon &\sim& \frac{1}{\sqrt{-W_0(-2ev^2)}}\\
  &\approx& \frac{1}{\sqrt{\ln\left(1/2ev^2\right)}}\label{eq:approximateHH}
\end{eqnarray}
in accordance to the previously published result of Hayakawa and Kuninaka \cite{HayakawaKuninaka:2004}. However, for the approximation \eqref{eq:approximateHH} to be useful the expression $\ln(1/2ev^2)$ has to approximate the Lambert $W$-function to a reasonable accuracy $\delta$. To this end we have to require 
\begin{eqnarray}
  \frac{1}{\sqrt{\ln\left(1/2ev^2\right)}} &>& \frac{1}{\delta}\left(\frac{1}{\sqrt{\ln\left(1/2ev^2\right)}}\right)^3\\
  \ln\left(1/2ev^2\right) &>& \frac{1}{\delta}\\
  v &<& e^{1/2\delta-0.5}
\end{eqnarray}
The critical velocity becomes exponentially small for increasing accuracy requirements. This severely limits its applicability. The full solution (to linear order in $\alpha$) Eq. \eqref{eq:firstorderepsilon} does not suffer from this limitation. 

The energy balance method unfortunately doesn't allow the calculation of higher order contributions to the energy loss, since this would require detailed knowledge of the trajectory. However, there is a method to calculate the contribution proportional to $\alpha^2$ which will be described in the next section.

\section{Consistency Method}\label{sec:consistency}

\noindent
In the previous section we have seen that the dominant term of the restitution coefficient is linear in the damping coefficient $\alpha$. With the help of a novel method it shall be attempted to improve the result to the next order in $\alpha$. It will be shown that for any viscoelastic problem of form \eqref{eq:dissforceprincipalstructure} the second order contribution to the coefficient of restitution is completely determined by the first order contribution. We assume that the next largest contribution is quadratic in $\alpha$, or generally that the restitution coefficient $\epsilon$ can be expressed as a power series in $\alpha$:
\begin{equation}
  \label{eq:epsilonassumoverorders}
  \epsilon(v)=\sum_{k=0}^{\infty}\alpha^kf_k(v)
\end{equation}
Note that the functions $f_k$ do not depend on $\alpha$. The zeroth function $f_0$ is 1 since the restitution coefficient has to be one for undamped collisions ($\alpha=0$). After the collision the relative velocity of the two particles will be 
\begin{equation}
  \label{eq:resultdirect}
  v^{\prime}=\epsilon(v)v~.
\end{equation}
Now we perform a time inversion, i.e. we start with the velocity $v^{\prime}$ at time $t_c$ and let the time run backwards until at time $t=0$ we arrive again at the initial velocity of the direct collision $v$. The equation of motion of this inverse problem reads
\begin{equation}
  \label{eq:eqnofmotioninverse}
  \ddot{x}+\phi-\alpha\frac{d\phi}{dx}=0
\end{equation}
Comparing with the equation of motion of the direct (forward time) collision (\ref{eq:eqnofmotion}) one notes that the only difference is the sign of the damping coefficient $\alpha$. In the inverse problem we have a negative damping $-\alpha$ so, in accordance with intuition, the inverse collision is an accelerated one. We now know the restitution coefficient of the inverse problem
\begin{equation}
  \label{eq:epsiloninvassumoverorders}
  \epsilon_{\rm inv}(v)=\sum\limits_{k=0}^{\infty}(-\alpha)^kf_k(v)~,
\end{equation}
again the only difference being the opposite sign of $\alpha$. As said before, starting the inverse collision at velocity $v^{\prime}$ we must arrive at the velocity $v$:
\begin{equation}
  \label{eq:resultinverse}
  v=\epsilon_{\rm inv}\left(v^{\prime}\right)v^{\prime}~.
\end{equation}
Inserting $v^\prime$ from Eq. \eqref{eq:resultdirect} into \eqref{eq:resultinverse} we obtain the consistency equation
\begin{equation}
  \label{eq:consistencyequation}
  \epsilon_{\rm inv}\left[v\epsilon(v)\right]\epsilon(v)=1~.
\end{equation}
Eq. \eqref{eq:consistencyequation}  has to be fulfilled for all restitution laws of form Eq. \eqref{eq:epsilonassumoverorders} regardless of the internal contact mechanism. Of course, different contact laws yield different velocity dependences for $\epsilon(v)$ and $\epsilon_{\rm inv}(v)$, however, the consistency requirement Eq. \eqref{eq:consistencyequation} in it's general form remains unchanged. Eq. \eqref{eq:consistencyequation} allows us to calculate the second order correction to the first order result Eq. \eqref{eq:firstorderepsilon}. We insert Eqs. \eqref{eq:epsilonassumoverorders} and \eqref{eq:resultinverse} into Eq. \eqref{eq:consistencyequation} using $f_0=1$ and
\begin{eqnarray}
  f_k\left[v\epsilon(v)\right]&=&f_k\left[v+\sum\limits_{k=1}^{\infty}\alpha^kvf_k(v)\right]\\
  &=&f_k(v)+\frac{\md f_k}{\md v}(v)\sum\limits_{k=1}^{\infty}\alpha^kvf_k(v) + \frac{1}{2}\frac{\md^2f_k}{\md v^2}\left(\sum\limits_{k=1}^{\infty}\alpha^kvf_k(v)\right)^2+\dots
\end{eqnarray}
and find 
\begin{eqnarray}
  1&=&1-\alpha^2\left(vf_1(v)\frac{\md f_1}{\md v}(v)+f_1^2(v)\right)+2\alpha^2f_2(v)+{\cal O}(\alpha^3)~.\label{eq:potenzreihe}
\end{eqnarray}
All terms on the right hand side of linear or higher order in $\alpha$ have to be zero which gives us the final expression for the second order term $f_2$
\begin{equation}
  f_2=\frac{vf_1f_1^{\prime}+f_1^2}{2}~,
\end{equation}
where $f_1^{'}$ means derivative with respect to the velocity $v$. When computing higher orders one notes that, unfortunately, terms of odd order in $\alpha$ vanish identically. We cannot, therefore, compute terms of $\varepsilon$ of order higher than $\alpha^2$ since half of the necessary equations are missing. From Eq. \eqref{eq:firstorderepsilon} we have 
\begin{equation}
  f_1(v)=-\sqrt{8}\sum\limits_{k=0}^{\infty}\frac{c_k}{\left[-W_0(-2ev^2)\right]^{k+1/2}}~.
\end{equation}
In order to calculate the first derivative of this expression with respect to $v$ we use 
\begin{equation}
  \frac{\md W_0(x)}{\md x}=\frac{W_0(x)}{x(1+W_0(x))}
\end{equation}
and find 
\begin{equation}
  f_1^{\prime}=\frac{\sqrt{32}}{v\left[1+W_0(-2ev^2)\right]}\sum\limits_{k=0}^\infty\frac{c_k\left(k+\frac{1}{2}\right)}{\left[-W_0(-2ev^2)\right]^{k+1/2}}~.
\end{equation}
With this expression we can calculate $vf_1f_1^{\prime}$
\begin{equation}
  vf_1f_1^{\prime}=-\frac{16}{1+W_0(-2ev^2)}\sum\limits_{k=0}^\infty\frac{1}{\left[-W_0(-2ev^2)\right]^{k+1}}\sum\limits_{i=0}^k\left(k-i+\frac{1}{2}\right)c_ic_{k-i}~.
\end{equation}
Finally $f_1^2$ reads
\begin{equation}
  f_1^2=8\sum\limits_{k=0}^\infty\frac{1}{\left[-W_0(-2ev^2)\right]^{k+1}}\sum\limits_{i=0}^kc_ic_{k-i}~.
\end{equation}
$f_2$ can be expressed in the form
\begin{equation}
  f_2(v)=\frac{4}{1+W_0(-2ev^2)}\sum_{k=0}^{\infty}\frac{d_k}{\left[-W_0(-2ev^2)\right]^k}~,
\end{equation}
where the $d_k$ are purely numerical constants wich can be computed from the $c_k$. 
\begin{eqnarray}
  d_0&=&-c_0^2\\
  d_k&=&-2\sum\limits_{i=0}^{k-1}(k-i-1)c_ic_{k-i-1}-\sum\limits_{i=0}^kc_ic_{k-i}~~~~\mbox{for}~~k>0~.
\end{eqnarray}
The first values of $c_k$ and $d_k$ are given in Table \ref{tab:firstvaluesckdk}. For $k=0$ and $k=1$ analytical expressions for both $c_k$ and $d_k$ can be found, the other constants have to be determined numerically. 
\begin{table}[htbp]
  \centerline{\begin{tabular}{|r|l|l|}
      \hline
      $k$ & $c_k$ & $d_k$\\
      \hline
      0&$\displaystyle\frac{\pi}{4}$ & $-\displaystyle\frac{\pi^2}{16}$\\[0.2cm]
      1&$\displaystyle\frac{\pi}{8}(1-\ln 4)$ & $\displaystyle-\frac{\pi^2}{16}(1-\ln 4)$\\[0.2cm]
      2&$-0.02237433$ & $0.250419$\\
      3&$-0.0076646$ & $0.029518$\\
      \hline
    \end{tabular}}
  \caption{First values for $c_k$ and $d_k$ (explanation given in text).}
  \label{tab:firstvaluesckdk}
\end{table}

\begin{figure}[htbp]
  \centerline{
    \includegraphics[height=5cm,clip]{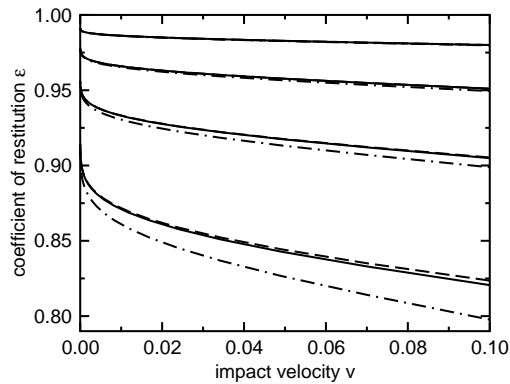}
  }
  \caption{Coefficient of restitution of two colliding disks for various values of $\alpha$. The full lines show the result of the numerical integration of Eq. (\ref{eq:eqnofmotion}). The dashed lines are the analytical solution up to second order in $\alpha$ (see Eq. \eqref{eq:finalsolution}). From top to botton the values of $\alpha$ are $0.02$, $0.05$, $0.1$ and $0.2$. To demonstrate the effect of the second order contribution in $\alpha$ the first order result is shown with dash-dotted line. One notes a large difference to the numerical result.}
  \label{fig:epsilondisk}
\end{figure}

\noindent We now have the final expression for the coefficient of normal restitution up to second order in $\alpha$:
\begin{equation}
  \epsilon=1-\sqrt{8}\,\alpha\sum\limits_{k=0}^\infty\frac{c_k}{\left[-W_0(-2ev^2)\right]^{k+1/2}} + \frac{4\alpha^2}{1+W_0(-2ev^2)}\sum\limits_{k=0}^\infty\frac{d_k}{\left[-W_0(-2ev^2)\right]^k}~.
  \label{eq:finalsolution}
\end{equation}
To check the analytical result the equation of motion Eq. (\ref{eq:eqnofmotion}) has been integrated numerically for different values of the scaled impact velocity $v$. The result for $\alpha=0.1$ is shown in Fig. \ref{fig:epsilondisk} (full line). Both analytical and numerical solution agree very well. 

\section{Conclusions}

\noindent In the present article the velocity dependence of the coefficient of normal restitution of colliding identical disks was derived. It turns out that it shows a significant dependence on velocity. It approaches one (no damping) for small velocities and decreases for increasing velocities. Consequently, for freely cooling systems it is problematic to use colliding disks as an example for particles with constant restitution coefficient. However, for driven systems the assumption of constant restitution may remain justified. For the computation of the coefficient of restitution it is not necessary to know exact details about the trajectories of the particles during the collision but instead a simple energy balance method together with a consistency consideration is sufficient to derive an analytical expression which is correct up to the second order of the damping parameter. It can be expected that the kinetic properties of gases of colliding disks differ significantly from the properties of gases of particles colliding with constant restitution coefficient. The analytic results for $\varepsilon(v)$ have been compared with results obtained by numerically integrating the equation of motion of the collision problem. Both solutions are in good agreement. 

The author wants to thank Thorsten P{\"o}schel and Nikolai Brilliantov for helpful discussions. 

\begin{appendix}
\section{Lambert $W$-function}\label{appendix:lambert}
\begin{figure}[htbp]
  \centerline{\includegraphics[width=7cm]{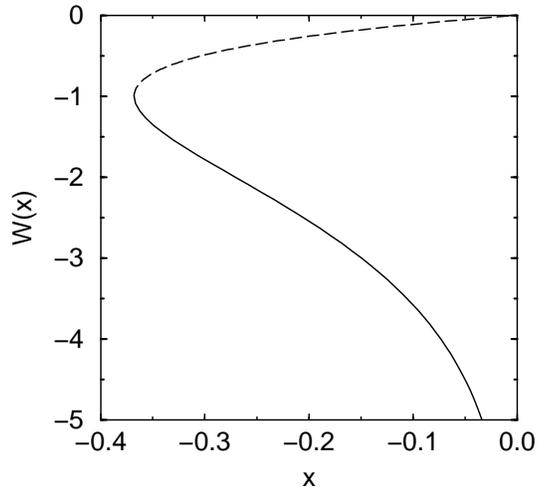}}
  \caption{Lambert $W$-function for negative argument. The dashed line is the (irrelevant) branch which approaches zero for $x\to0$, the full line is the branch which displays the required behaviour.}
  \label{fig:lambert}
\end{figure}

\noindent
The Lambert $W$-function of argument $x$ is implicitely defined by the following equation \cite{Corless:1996}
\begin{equation}
  \label{eq:lambertdef}
  We^W=x~.
\end{equation}
The defining equation has one real solution for positive values of $x$ but two for negative $x$. Fig. \ref{fig:lambert} shows a plot of this function for the relevant case $x<0$. 

One solution approaches zero like $W(x)\to x$ for $x\to 0$, the other approaches $-\infty$. It is clear that the first solution ($W(x)\to x$) is unphysical since it would yield a force $\phi\to1$ as $x\to0$, i.e., a nonvanishing contact force for zero compression. To avoid confusion between the two branches the relevant solution will be called $W_0$ instead of $W$. For small negative values of $x$ the function $W_0$ can be approximated by 
\begin{equation}
  W_0\approx-\ln\left(-\frac{1}{x}\right)-\ln\ln\left(-\frac{1}{x}\right)
\end{equation}
It shall be mentioned that there is no real $W_0$ for $x<-e^{-1}$ (see again Fig. \ref{fig:lambert}). This, however, is of no impact to our solution since the we are restricted to small arguments both in the force law Eq. \eqref{eq:scaledexplicitlaw} as well as in the final solution Eq. \eqref{eq:finalsolution}. More specifically, an argument $x=-e^{-1}$ in Eq. \eqref{eq:scaledexplicitlaw} would correspond to compressions comparable to the particle radius, in Eq. \eqref{eq:finalsolution} it would correspond to velocities comparable to the speed of sound in the material, both cases are clearly beyond the limits of viscoelastic approximation. 

The first derivative of the Lambert $W$-function with respect to its argument can be calculated by differentiating both sides of Eq. (\ref{eq:lambertdef}):
\begin{eqnarray}
  \frac{\md W}{\md x}e^W\left(1+W\right)&=&1\\
  \frac{\md W}{\md x}&=&\frac{1}{e^W(1+W)}
\end{eqnarray}
or using the defining equation again 
\begin{eqnarray}
  e^W&=&\frac{x}{W}\\
  \frac{\md W}{\md x}&=&\frac{W}{x(1+W)}
\end{eqnarray}
These expressions hold for both the real $W$-functions ($W$ and $W_0$). 
\end{appendix}

\end{document}